\documentclass[aip,reprint]{revtex4-1}
\usepackage{graphicx}
\usepackage{color}
\usepackage[caption=false]{subfig}
\usepackage{amsmath,amssymb}
\usepackage{float}
\usepackage[utf8]{inputenc}
\usepackage{mathptmx}
\usepackage{mathtools}

\begin{document}
\title{Nonlinear dynamics of emergent traveling waves in a reaction-Cattaneo system}
\author{Pushpita Ghosh}\email{pghosh@tifrh.res.in}
\affiliation{Tata Institute of Fundamental Research Hyderabad,  Telangana 500046, India}
\author{Deb Shankar Ray}
\affiliation{Indian Association for the Cultivation of Science, West Bengal 700032, India.}

\begin{abstract}
Standard diffusion equation is based on Brownian motion of the dispersing species without considering persistence in the movement of the individuals. This description allows for the instantaneous spreading of the transported species over an arbitrarily large distances from their original location predicting infinite velocities. This feature is unrealistic particularly while considering biological invasion dynamics and a better description needs the consideration of dispersal with inertia. We here examine the behavior of non-infinitesimal perturbation on the steady state of an one-dimensional reaction-Cattaneo system with a cubic polynomial source term describing population dynamics or flame propagation models. It has been shown analytically that while linear analysis predicts stability of the homogeneous state, consideration of nonlinear contribution leads to a growth of spatiotemporal perturbation as a traveling wave. We show that the presence of a small finite relaxation time of the diffusive flux modifies the speed of the traveling wave. Specifically, we find that the wave speed decays with an increase of a finite relaxation time of flux. Our analytical predictions are well corroborated with the numerical results.
\end{abstract}
\maketitle
\newpage

\section{Introduction}

Reaction-transport models are often conveniently  used to describe the spatiotemporal dynamics of systems in a wide variety of fields including biology, ecology, chemistry and physics ~\cite{Murray,Epstein,Cross}. Macroscopically, interactions among the individual species and with the surrounding environment are described by a reaction kinetics or source term while the dispersal is typically modeled by diffusive flux term. The interplay of dispersal and source kinetics is responsible for the occurrence of a rich variety of spatiotemporal instabilities such as stationary spatial patterns and propagating waves across many reaction-diffusion systems~\cite{Murray,Epstein,Cross,Turing}. However, standard diffusion equation, which is based on Brownian motion of the constituting  particles of a system, fails to offer a satisfactory description of transport in many cases when the motion is persistent or in other words, when the correlation between successive motions of dispersive particle is large~\cite{Horsthemke1999,Hillen,Fedotov}.  This results in instantaneous spreading of particles from their original location over a infinitely large distances and consequently predicts infinite velocities of propagation of the diffusing particles. This features are not entirely realistic and could yield inaccurate or simply unphysical results in many situations of front propagation particularly in biology~\cite{Holmes,Stinner,Codling}.  There are several instances, where we find signature of wave propagation in biology, e.g, invasions of populations of bacteria and viruses, growth of tumors, wake of advantageous genes, propagation of epidemics, etc. Practically, all organisms disperse at a finite velocity with some persistence in motion at least over short time intervals.

This pathological pitfall of the diffusion equation stems from not taking into account the inertia of the Brownian particles and therefore can be eliminated through invoking a nonzero relaxation time of the flux following Cattaneo's modification of the original Fick's law of diffusion~\cite{Cattaneo}. The delay in the flux equation may be perceived as persistent random walk valid from ballistic limit to diffusive limit so that one can associate a finite speed of propagation of the dispersing species. The reaction-Cattaneo system and its variants have been investigated by a number of authors in several contexts~\cite{ABDUSALAM2006,PGhosh2009,PGhosh2010,Tilles2019,Vergni,Serjei,Sabelnikov,MISHRA2012}. Here, our aim is to study the nonlinear dynamics of spatially-extended system focusing on the phenomenon of propagating fronts or traveling waves following a generalized reaction-Cattaneo equation. Traveling waves or wave fronts are solution of spatially-extended dynamical systems having multiple steady states. The first step to characterize the nature of stability of such steady states of a dynamical system primarily rests on linear stability analysis by following
the dynamics of temporal and spatiotemporal perturbations
in the close neighborhood of the homogenous steady
states~\cite{Strogatz}. It is thus imperative that the evolution of infinitesimal perturbation
when applied on a homogeneous steady state of the system is
well-described by linear equations with exponential solutions. 
The nature of the steady state remains predictable within a time-scale
characteristic of the inverse of rate constant associated with the
exponential solutions. However, when the perturbation is
finite and large the linear equations need to be supplemented by
nonlinear contributions either fully or partially with appropriate
convergence. The resulting evolution of non-infinitesimal
perturbation may then capture a number of essential nonlinear
features of the concerned dynamical system, which otherwise clearly
remain outside the scope of linear analysis. The present paper
addresses such an issue in the context of generation of traveling
waves under finite perturbation on a homogeneous stable steady state
by taking care of the terms of all orders of the polynomial source function in a reaction-Cattaneo system, rendering the solution to be exact.

The behavior of noninfinitesimal perturbation has been the subject interest of
in several contexts \cite{Dressler,Torcini,Aurell,SKar,Cencini,SDutta,Politi}.
For example, the notion of Lyapunov exponents,
which usually characterize the average local stability properties of
a system to describe the rate at which small volumes expand or
contract in different directions, has been generalized
\cite{Dressler} to higher order derivatives. These derivatives
effectively take care of nonlinear distortions. Nonlinear mechanisms
play a key role in controlling the evolution of the localized
distributions in extended chaotic system \cite{Torcini}. The effects
of finite perturbations in fully developed turbulence have been
discussed by introducing a measure of chaotically degree associated
with the velocity field \cite{Aurell}. Interplay of nonlinear
diffusion and self-limiting growth process has been analyzed to
study the rate of spread of wave front in reaction-diffusion models
\cite{SKar}. By employing coupled map lattices to mimic spatially
extended chaotic systems finite and infinitesimal amplitude
perturbation evolution has been investigated \cite{Cencini} to
characterize information transport. Nonlinear perturbation has been
found to be useful \cite{SDutta} in deriving instability conditions for
pattern formation induced by additive noise and in analyzing the
type of patterns \cite{PGhosh2009,PGhosh2009_2} using Galerkin techniques.

        The above consideration suggests that nonlinear dynamics of
non-infinitesimal perturbation may be generically different from what
is expected from linear analysis. Two points in this context are in
order. First, traditionally a perturbation series in general, is
susceptible to the problem of convergence when the perturbation is
truly non-infinitesimal. For a polynomial source function as
employed in the present problem  which forces the higher order
perturbation terms to remain finite in number, it is possible in some cases to
keep track of all the terms for nonlinear analysis. Second, since
the coefficients of expansion of the source function characterize
the nature of the homogeneous steady state, they play a significant
role in dynamical evolution of non-infinitesimal perturbation. It is
therefore worthwhile to explore this dynamics in spatially extended
systems with a polynomial source function. To this end a full scale
nonlinear stability analysis has been performed recently on a
reaction-transport system with a cubic polynomial source function to
show~\cite{PGhosh2009,PGhosh2010} how a finite perturbation on a homogeneous steady state may lead to spatiotemporal instability. Such systems may also
exhibit~\cite{SSen} sec-hyperbolic instability in the stationary
limit and an inhomogeneous profile of reaction component reminiscent
of a solitary wave. The present paper is an attempt in this
direction pertaining to wave propagation of a front. We consider a system of individuals that move in a correlated way with a finite speed, and that reproduce and die with prescribed reaction kinetics. Specifically, we consider an one-component reaction-Cattaneo system with a cubic polynomial source term describing e.g., population dynamics of microorganisms or flame propagation. The evolution equation for perturbation which contains nonlinearity up to third order is studied in the phase plane in search of a traveling wave solution. An interesting relaxation time induced modification of traveling wave speed is observed. Specifically our object is three-fold:

($\emph{i}$) to explain how a non-infinitesimal spatiotemporal perturbation on a
homogeneous linearly stable steady state grows as a traveling wave with
finite speed. Or, in other words, how the finite perturbation induces instability and what factors other than the diffusion coefficient control the speed of the wave front.

($\emph{ii}$) to understand whether the finite relaxation time of the flux
modifies the growth of a traveling wave and what determines the
associated wave speed for the reaction-Cattaneo system.

($\emph{iii}$) to explore a generic situation that concerns a linearly
stable state which turns out to be unstable in a nonlinear analysis
due to the characteristic nature of the expansion coefficients of the
reaction term.

The organization of the paper is as follows; in Sec II we introduce
the reaction-Cattaneo system with a cubic polynomial source term
which describes population dynamics with strong Allee effect or flame propagation and set up the  nonlinear equation for spatiotemporal evolution of
the perturbation around a homogeneous stable state. Traveling wave
transformation is then used to demonstrate the growth of finite
perturbation as traveling wave with a finite speed. The explicit
traveling wave solution and the wave speed are derived to expose the
relaxation effect. Sec-III, describes the results of numerical simulations of the model system. The paper is concluded in Sec IV.

\section{Non-infinitesimal perturbation and the Growth of a traveling wave}

Macroscopically the dynamics of a field variable $u(x,t)$, can be described from the equation of continuity with a source or reaction term $f(u)$
\begin{eqnarray}\label {eq1}
\frac{\partial u(x,t)}{\partial t}=-\frac{\partial J(x,t)}{\partial x}+f(u) 
 \end{eqnarray}
where $J(x,t)$ is the flux of $u(x,t)$ and can be derived from Fick's law of diffusion as follows
\begin{eqnarray}\label {eq2}
J(x,t)=-D \frac{\partial u(x,t)}{\partial x}.
\end{eqnarray}
where $D$ is the diffusion coefficient of the species $u$.
The diffusion equation that lies at the heart of the treatment suffers from one
inherent pitfall. Essentially this is due to the neglect of inertia
in the Brownian dynamics of the reacting species. The Gaussian form
of the solution of diffusion equation with the initial condition
that at $ t=0 $ all the particles reside at $x=0$ suggests that the
particles can be traced at a very very large distance from the
initial location at $ x=0 $ even when the time is very short. This
implies that the particles have infinite speed. A way to bypass this
difficulty is to introduce the finite relaxation time of the
diffusive flux. Following Cattaneo's modification of Fick's law of
the form
\begin{eqnarray}\label{eq3}
J(x,t+\tau) = -D\frac{\partial u(x,t)}{\partial x}
\end{eqnarray}
one takes care of the adjustment of the concentration gradient at
time $t$ to a delayed flux at a later time $t+\tau$. Expanding $J$
in Eq.(\ref{eq3}) up to first order in $\tau$ in a Taylor series and
differentiating the resulting equation with respect to $x$ we obtain
\begin{eqnarray}\label{eq4}
\frac{\partial J(x,t)}{\partial x} + \tau \frac{\partial^{2}
J(x,t)}{\partial x \partial t} =
-D\frac{\partial^{2}u(x,t)}{\partial x^{2}}
\end{eqnarray}
Differentiation of Eq.(\ref{eq1}) with respect to $t$ followed by elimination of $J(x,t)$ from the
resulting equation and Eq.(\ref{eq4}) leads us to the following reaction-Cattaneo's equation
\begin{eqnarray}\label{eq5}
\tau \frac{\partial^{2} u(x,t)}{\partial t^{2}} + [1-\tau
f'(u)]\frac{\partial u(x,t)}{\partial t} =  D \frac{\partial^{2}
u(x,t)}{\partial x^{2}} +  f(u)
\end{eqnarray}
An interesting aspect of this equation is that this modification of
diffusive motion may be perceived as a persistent random walk valid
from ballistic limit to diffusive limit so that one can associate a
finite speed of propagation of the dispersing species. 

We now specialize the source function by a cubic polynomial, represented by $f(u)=-u^{3}+3u^{2}-u$, reminiscent of flame propagation or population dynamics models. Cubic polynomial source functions are widely known for many reaction-diffusion problems~\cite{Epstein,Scott}, e. g., in flame propagation~\cite{Zeldovich}, in impulse propagation along active nerve fibre ~\cite{Offner,Zemskov}, in autocatalytic reactions comprising chlorite-iodide and arsenite-iodide reactions~\cite{Kepper,SSen2010,PGhosh2019} and population dynamics with strong Allee effect~\cite{Taylor,Johnson2006}. Populations described using Allee effect deviates from the logistic growth kinetics and models exhibit more complex and subtle dynamics, including reduced growth at low densities~\cite{Oikos,Neufeld,Johnson2006} and extinction below a critical density threshold ~\cite{Courchamp,Taylor}.

To understand the spatiotemporal evolution of density of a dynamical system described by aforementioned source kinetics, 
we begin by considering the homogeneous steady states which are the fixed points $u_{0}$, of the
dynamical system and obtained as solutions of the algebraic equation
\begin{eqnarray}
f(u_{0})=0 
\end{eqnarray}
The spatiotemporal perturbation on a homogeneous steady state is
given by
\begin{eqnarray}\label{eq6}
\delta u(x,t) = u(x,t) - u_{0} 
\end{eqnarray}
We first assume that the homogeneous steady state $u_{0}$ is
linearly stable with respect to infinitesimal perturbation so that
$f'(u_{0})$ is negative. Since the perturbation (\ref{eq6}) is finite, a
nonlinear analysis of the problem requires all the derivatives of
the source term beyond $f'(u_{0})$ around the homogeneous linearly
stable steady state. 
Since the source or kinetic term $f(u)$ is a cubic polynomial in
form , the derivatives higher than third are all zero.

To this end we begin with the following two expansions for $f(u)$ and $f'(u)$ for
cubic polynomial function.
\begin{eqnarray}
f(u) &=& f(u_{0} + \delta u) \nonumber  \\
&= &f(u_{0}) + f'(u_{0})(\delta u) +
\frac{f''(u_{0})}{2}(\delta u)^{2} + \frac{f'''(u_{0})}{6}(\delta
u)^{3} \label{eq7}
\\ 
f'(u)&=&f'(u_{0} + \delta u)\nonumber  \\
&=&f'(u_{0}) + f''(u_{0})(\delta u) +
\frac{f'''(u_{0})}{2}(\delta u)^{2} \label{eq8}
\end{eqnarray}
Making use of the expressions (\ref{eq7}) and (\ref{eq8}) in Eq.(\ref{eq5}) and writing $u =
u_{0} + \delta u(x,t)$ we obtain
\begin{widetext}
\begin{eqnarray}\label{eq9}
\tau  \frac{\partial^{2}}{\partial t^{2}}(\delta u) + \left[ 1 -
\tau \left\{ f'(u_{0})+f''(u_{0})(\delta u) +
\frac{f'''(u_{0})}{2}(\delta u)^{2} \right\} \right]
\frac{\partial}{\partial t}(\delta u)  = D
\frac{\partial^{2}}{\partial x^{2}}(\delta u) +  \left[f'(u_{0}) +
\frac{f''(u_{0})}{2}(\delta u)^{2} + \frac{f'''(u_{0})}{6}(\delta
u)^{3}\right] \nonumber \\ 
\end{eqnarray}
\end{widetext}

Eq.(\ref{eq9}) is the starting point of our further analysis. We begin with
a traveling wave transformation by writing
\begin{eqnarray}
z = x - vt  \label{eq10}
\end{eqnarray}
where $v$ is the velocity of the wave associated with the
perturbation  $\delta u(x,t)\equiv  \xi(z)$. $v$ is to be determined.

With the traveling wave transformation (\ref{eq10}) and abbreviation
$\delta u(x,t)\equiv \xi(z)$ as before, Eq.(\ref{eq9}) becomes
\begin{eqnarray}\label{eq11}
\tau v^{2}\frac{\partial^{2} \xi}{\partial z^{2}} &-& \left[ 1 -
\tau \left\{ f'(u_{0})+f''(u_{0})\xi + \frac{f'''(u_{0})}{2}\xi^{2}
\right\} \right]v\frac{\partial \xi}{\partial z} \nonumber \\
&=& D \frac{\partial^{2} \xi}{\partial z^{2}} + \left[f'(u_{0}) +
\frac{f''(u_{0})}{2}\xi^{2} + \frac{f'''(u_{0})}{6}\xi^{3}\right]
\end{eqnarray}
The above equation may be rearranged to obtain
\begin{widetext}
\begin{eqnarray}\label{eq12}
(D  - \tau v^{2})\frac{\partial^{2} \xi}{\partial z^{2}} + v \left[
1 - \tau \left\{ f'(u_{0})+f''(u_{0})\xi +
\frac{f'''(u_{0})}{2}\xi^{2} \right\} \right]\frac{\partial
\xi}{\partial z} =  L(\xi)
\end{eqnarray}
\end{widetext}

The right hand side of Eq.(\ref{eq12}) can now be expressed as
\begin{eqnarray}\label{eq13}
L(\xi) = \xi (\gamma \xi - \alpha)(\xi - \beta)=\gamma \xi^{3} -
(\alpha + \beta \gamma)\xi^{2} + \alpha \beta \xi \nonumber \\
\end{eqnarray}
where $\alpha$, $\beta$ and $\gamma$ are three constants to be determined as follows. 
Eq.(\ref{eq12}) forms a dynamical system for perturbation variable
$\xi(z)$. 
$L(\xi)$ is a cubic polynomial whose explicit form is to
be determined by the characteristics  of the cubic polynomial source
term $f(u)$ of the original dynamical system (\ref{eq5}).
Comparison between the right hand sides of Eq.(\ref{eq12}) and Eq.(\ref{eq13})
yields
\begin{eqnarray}\label{eq14}
 \gamma=-\frac{f'''(u_{0})}{6}, \quad \alpha + \beta \gamma = \frac{f''(u_{0})}{2}, \quad \alpha \beta = -
 f'(u_{0})
\end{eqnarray}

To explore the nature of dynamical system described by Eq.(\ref{eq12}) we
first note the three fixed points determined by the zeros of
$L(\xi)$, i. e., $\xi_{0}=0$, $\xi_{0}=\alpha/\gamma$ and
$\xi_{0}=\beta$. Now for the homogeneous stable steady state of the
original dynamical system Eq.(\ref{eq5}) $f'(u_{0}) <  0 $. This implies
$\alpha \beta > 0 $ (see Eq.(\ref{eq14})) and consequently from the
positivity of
\begin{center}
$L'(\xi_{0}) = 3\gamma \xi_{0}^{2}  - 2(\alpha + \beta)\xi_{0} +
\alpha\beta$
\end{center}
at $\xi_{0}=0$, we may infer that $\xi_{0} = 0$ is an unstable
steady state of the dynamical system (\ref{eq13}) for the  perturbation
variable. We may anticipate that a traveling wave may emanate from
this point. The assumption of a traveling wave form of solution thus
converts a partial differential equation (\ref{eq9}) into an ordinary
differential equation (\ref{eq12}) which is equivalent to the following two
 first order systems 
\begin{eqnarray}\label{eq15}
 \frac{d \xi}{d z} &=& w \nonumber
\\
\frac{d w}{d z} &=& \dfrac{\xi(\gamma \xi - \alpha)(\xi - \beta)}{D
 - \tau v^{2}}\nonumber \\ &-& \dfrac{v}{D - \tau v^{2}}\left[ 1 - \tau \left\{f' + f''
\xi + \frac{f'''}{2} \xi^{2}\right\} \right]w
\end{eqnarray}
for which the phase plane is described by $(\xi, w)$.
The wave speed $v$, an unknown parameter must be determined from the
analysis.

\begin{figure}
\includegraphics[width=5cm,origin=c,angle=0]{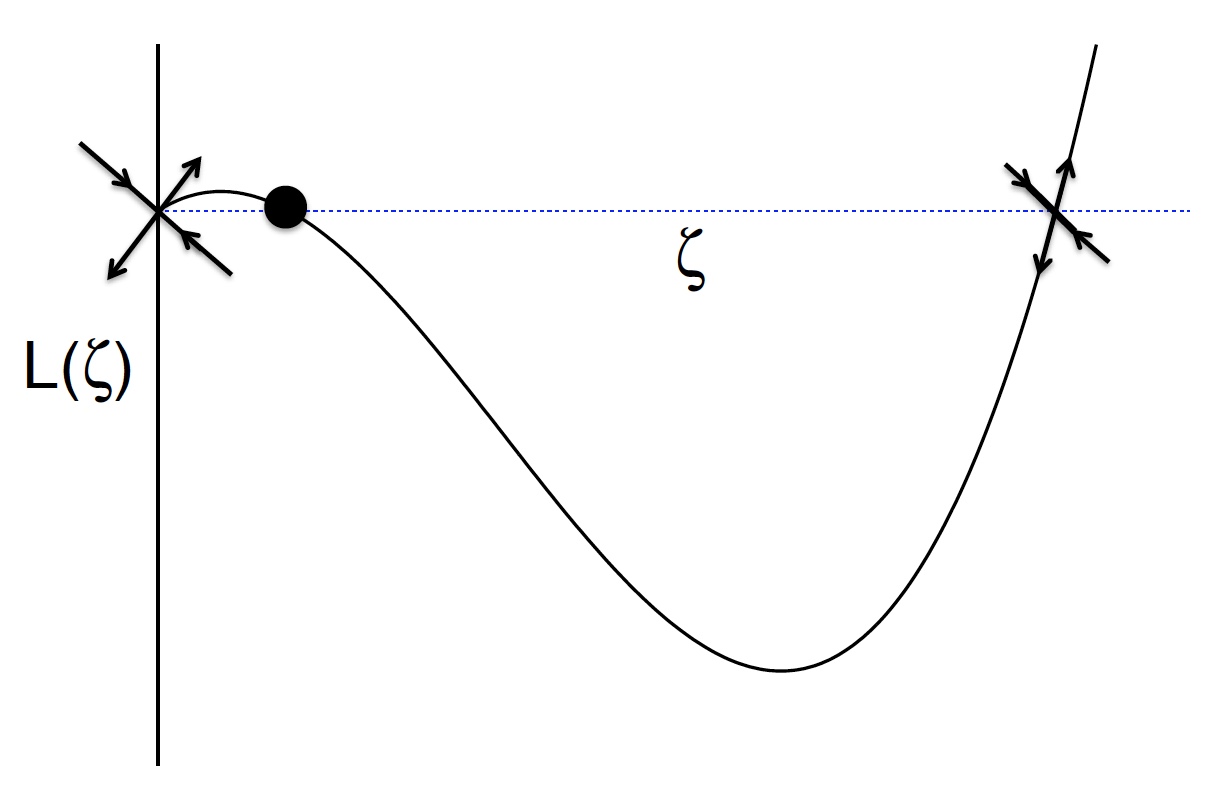}\caption{The schematic form of the cubic polynomial source
function and the three fixed points.}\label{cubic}
\end{figure}

\begin{figure}
\includegraphics[width=5cm,origin=c,angle=0]{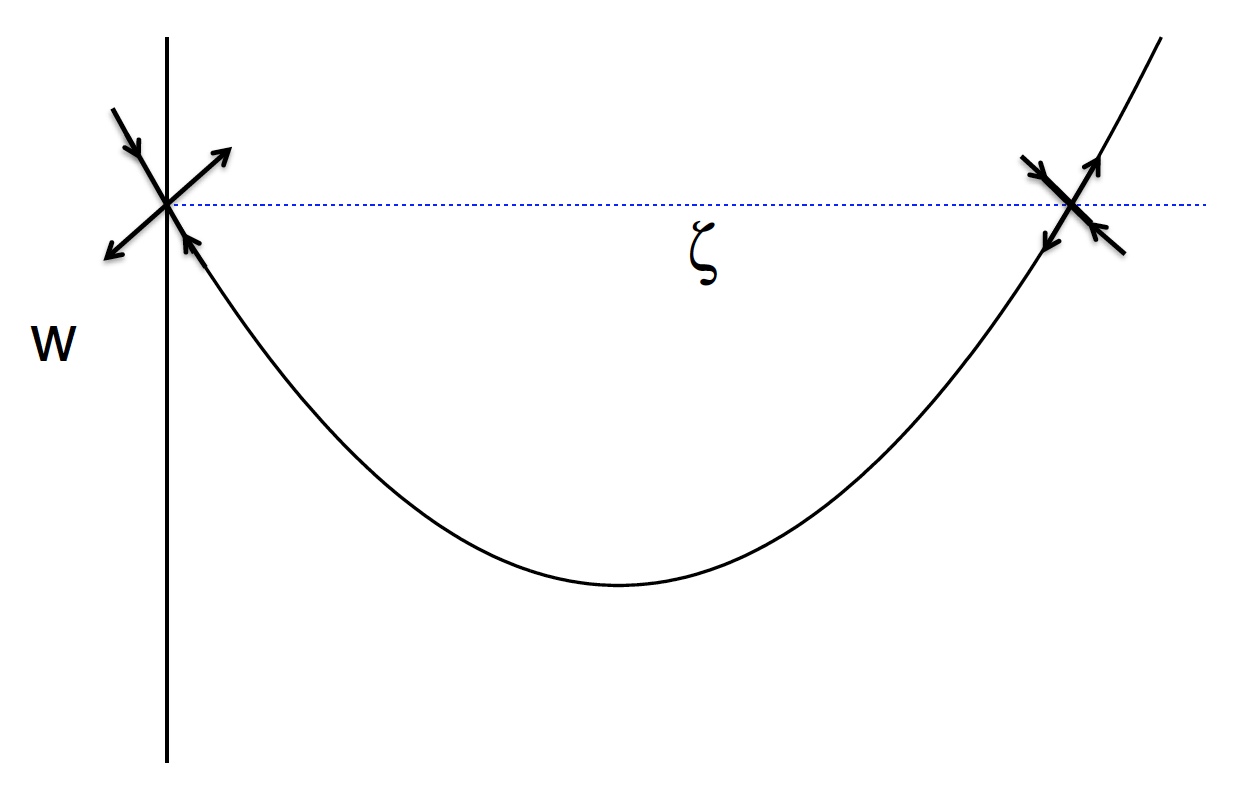}\caption{The schematic trial solution~(\ref{eq16}) connecting the two saddle points}\label{bistable}
\end{figure}

For the cubic shaped function $L(\xi)$ the system has
three fixed points at which the velocity vector $(\frac{d\xi}{dz},
\frac{dw}{dz})$ =$(0,0)$.
Three fixed points are $(0,0)$, $(\alpha/\gamma, 0)$ and $(\beta,
0)$. It is easy to show that the corresponding
fixed points are $(0,0)$, $(\frac{3-\sqrt{5}}{2},0)$ and
$(\frac{3+\sqrt{5}}{2},0)$ of which first and third are saddle
points while the other is a node. The function $L(\xi)$ and the fixed
points are shown in Fig.\ref{cubic}. The velocity vector spans a solution
trajectory which is an unique function of the position vector
$(\xi,w)$. Eq.(\ref{eq15}) can thus be studied in the $(\xi,w)$ phase
plane of Fig.\ref{bistable}.
To find a physically meaningful solution of Eq.(\ref{eq9}), it is necessary
that $(\xi,w)$ must remain bounded for large values of $z$. This
requirement can be satisfied by adjusting the traveling wave
velocity $v$ of the perturbation so that a trajectory leaving one of
the saddle at $ z = -\infty $ lies on a trajectory approaching the
other at $ z = + \infty $. Thus the traveling wave speed is fixed by
the condition that the solution and its derivative must remain
finite. The method of solution is quite standard \cite{Scott}.

Let a trial function or trajectory that intersects the two saddle
points is
\begin{eqnarray}\label{eq16}
w = K\xi(\gamma\xi - \alpha),
\end{eqnarray}
which is depicted in Fig.\ref{bistable} where $K$ is a constant to be determined from the known parameters of the source function and diffusion coefficient. 
Eq.(\ref{eq16}) implies
\begin{eqnarray}\label{eq17}
\frac{d w}{d\xi} = K(2\gamma\xi - \alpha)
\end{eqnarray}
From Eq.(\ref{eq15}) we immediately arrive at the following equation
\begin{eqnarray}\label{eq18}
\frac{d w}{d \xi} &=& \dfrac{\xi(\gamma \xi - \alpha)(\xi - \beta)}{(D
- \tau v^{2})w} \nonumber \\
&-& \dfrac{v}{(D  - \tau v^{2})}\left[ 1 - \tau
\left\{f' + f'' \xi +\frac{f'''}{2} \xi{2}\right\}\right] 
\end{eqnarray}

\begin{figure}
\includegraphics[width=8cm,origin=c,angle=0]{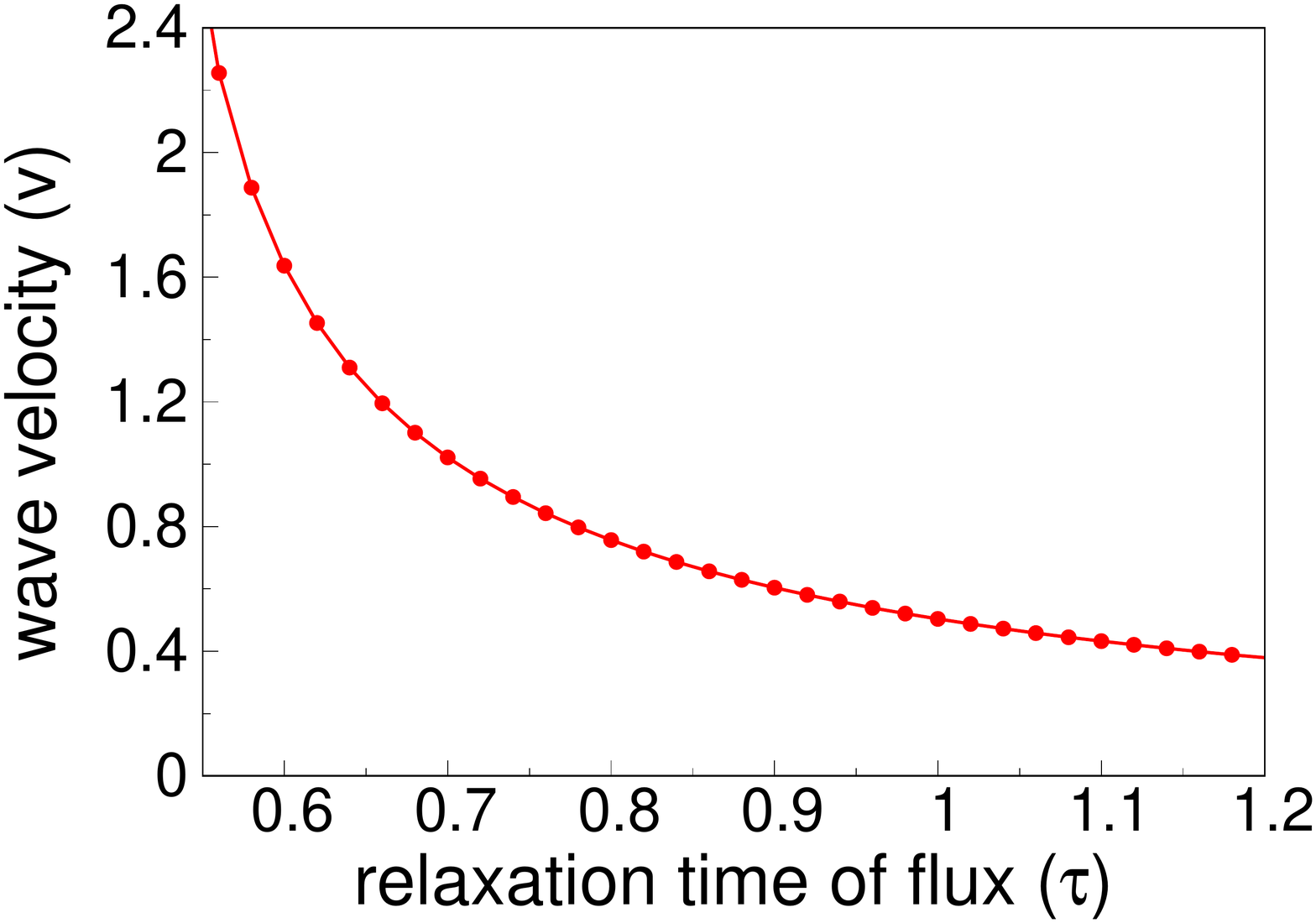}\caption{Plot of velocity of the traveling waves moving to the right for the set of parameters; $\alpha=2.612$, $\beta=0.381$, $\gamma=1.0$, $D=0.5$ and $c_{0}=1.0$ as a function of the relaxation time of the flux.}\label{velocity}
\end{figure}

Proceeding with the assumption that $\tau \frac{f'''(u_{0})}{2} \xi^{2}$ in Eq.(\ref{eq18}) is small, we
finally obtain
\begin{eqnarray}\label{eq19}
1 + \tau f'' v K &=& 2 \gamma K^{2} (D - \tau v^{2}) \quad\quad (a)
\nonumber
\\
\beta + v k (1 - \tau f') &=& \alpha K^{2} (D - \tau v^{2})
\quad\quad\;\;(b)
\end{eqnarray}
$K$ and $v$ can be determined from Eq.(\ref{eq19}). By systematic analysis, we obtained the expression where $K$ and $v$ is related by the following relation
\begin{equation}
Kv=\frac{(\alpha-2\gamma\beta)}{2(\gamma-\tau \alpha^2)}.\label{eq20}
\end{equation}
and the constant $K$ can be determined from the following expression 
\begin{eqnarray}\label{eq21}
2\gamma D K^2 &=& 1.0+\tau \frac{2(\alpha-2\gamma\beta)(\alpha+\gamma\beta)}{2(\gamma-\tau \alpha^2)}+\frac{\tau\gamma(\alpha-2\gamma\beta)^2}{2(\gamma-\tau\alpha^2)^2}\nonumber \\
&=& S
\end{eqnarray}
provided $(K^2=S/2\gamma D > 0)$. It turns out that the solution does not exist for a particular range of the relaxation time of flux ($0.226 < \tau < 0.502$). Therefore, the accepatable values of $\tau$ should be either $\tau<0.226$ or $\tau>0.502$. This outcome is one of the caveats of our analytical study. As we are mostly intereseted to study the hyperbolic regime when $\tau$ is not too small, in the present purpose, we consider a finite value of $\tau>0.502$ as determined from the analytical calculations for the rest of the study. 
As we proceed, $K$ and $v$ are obtained as follows;
\begin{eqnarray}\label{eq22}
K &=& \pm \left[\frac{S}{2\gamma D}\right]^{\frac{1}{2}} \\
v &=& \dfrac{(\alpha - 2 \gamma
\beta)}{2K(\gamma-\tau \alpha^2)}\label{eq23}
\end{eqnarray}
It is found from the expression (\ref{eq23}) that the traveling wave
velocity depends on the reactions parameters, diffusion coefficient as well as on the relaxation time of the flux. Fig.\ref{velocity} depicts the dependence of the velocity of the moving front on the relaxation time of the flux for $\tau > 0.502$ for negative values of $K$. For small but finite $\tau$, it predicts the decay in the traveling wave velocity with an increase of $\tau$ in a nonlinear fashion. However, for larger values of $\tau$, the decay is less.

\begin{figure}
\includegraphics[width=8cm,origin=c,angle=0]{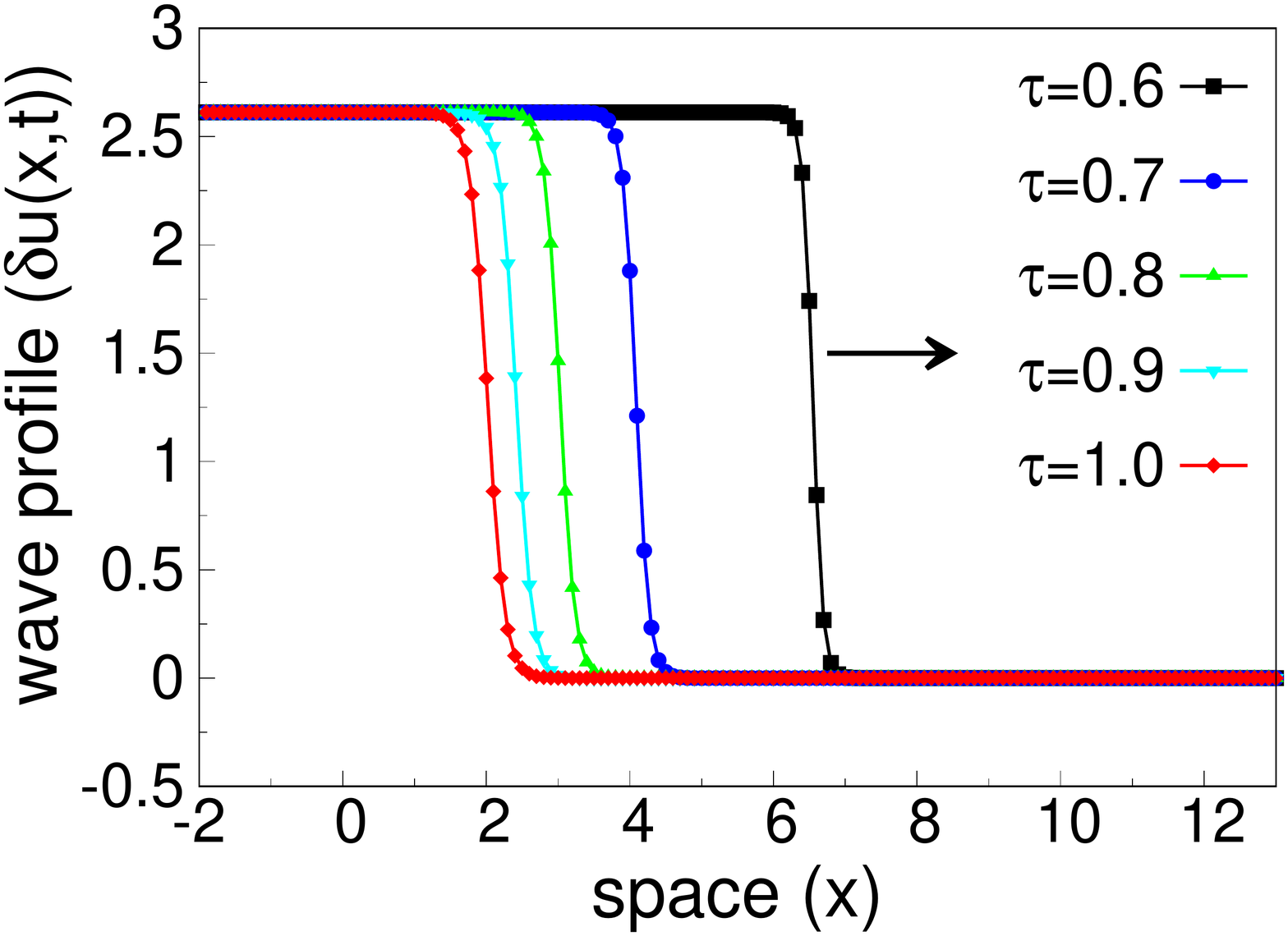}\caption{Density profiles of the traveling waves connecting the two steady states and moving to the right for the set of parameters; $\alpha=2.612$, $\beta=0.381$, $\gamma=1.0$, $D=0.5$ and $c_{0}=1.0$, for different delay times of the flux:
(a) $\tau=0.6$, (b) $\tau=0.7$, (c) $\tau=0.8$, (d) $\tau=0.9$, and  (d) $\tau=1.0$.}\label{wavefront}
\end{figure}

It is noteworthy that in the absence of the relaxation
effect, i. e., when $\tau=0$, the wave velocity reduces to the form $[v= \pm \sqrt{2\gamma D}(\frac{\alpha-2\gamma\beta}{2\gamma})]$
as obtained in case of no inertia $\tau=0$. Finally, from the first of Eq.(\ref{eq16}) and Eq.(\ref{eq17}) one arrives at
\begin{eqnarray}\label{eq24}
\frac{d\xi}{dz} = \pm K \xi (\gamma \xi -
\alpha)
\end{eqnarray}
which can be integrated to obtain the traveling wave solutions of
the partial differential equation (\ref{eq12}) as
\begin{eqnarray} \label{eq25}
\delta u(x,t) = \dfrac{\alpha}{\gamma +  c_{0}e^{\pm\frac{\alpha
(x-vt)}{ K}}}
\end{eqnarray}
where $c_{0}$ is a constant, comes from integration.
The typical wave forms are shown in Fig.\ref{wavefront} for several values of the relaxation time of flux.
It is apparent that small but finite relaxation time leads to a decay of speed of the traveling wave. This observation is the key result of the present analysis. It suggests that when the motion is correlated, we could see reduced speed of the propagating fronts at a particular time in presence of a small but finite $\tau$. Moreover, the shape of the wave front is also dependent on the value of $\tau$ as denoted by the traveling wave solution given in Eq.(\ref{eq25}).

The above solution clearly asserts that in contrast to an
infinitesimal perturbation which decays monotonically, a finite
perturbation on a homogeneous stable state may give rise to a
traveling wave solution $\delta u(x,t)$ denoted by the expression in Eq.(\ref{eq25})) with a finite speed $v$, as given in the expression~(\ref{eq23}). The solution and the speed are characterized not only by diffusion coefficient of the reacting species but also by the nature of nonlinearity of the
reaction term through $\alpha$, $\beta$, $\gamma$ and the presence of a small time delay in the flux. Since the nature
of the steady state characterizes these coefficients the nonlinear
analysis clearly reveals their signature in destabilization of the
linearly stable state. The traveling wave of the form (\ref{eq24}) is also
well known in Fisher equation and its variants in the context of
self-limiting growth models \cite{Murray}.

\section{Numerical results}
In order to compare the analytical predictions obtained from the aforesaid analytical study, we carry out numerical simulation of the spatially-extended reaction-transport model described by Eq.(\ref{eq5}) in the presence of a small but finite relaxation time of the flux. Using the method described in~\cite{Logan,Sabelnikov} the hyperbolic Eq.(\ref{eq5}) can be recast into the characteristic form given below
\begin{eqnarray}
\frac{\partial u_+}{\partial t} + \frac{1}{\sqrt{\tau}}\frac{\partial u_+}{\partial x} = -\frac{1}{2\tau}(u_+ - u_-)+\frac{f(u)}{2}\label{eq26}\\
\frac{\partial u_-}{\partial t} - \frac{1}{\sqrt{\tau}}\frac{\partial u_-}{\partial x} = +\frac{1}{2\tau}(u_+ - u_-)+\frac{f(u)}{2}\label{eq27}
\end{eqnarray}
where $u_+$ and $u_-$ are the Reimann invariants and defined as 
$u_+=\frac{1}{2}(u+j)$ and $u_-=\frac{1}{2}(u-j)$. Here $j$ stands for the flux of the system. In other words, the total density is divided into two subpopulations such that $u=(u_+ + u_-)$ and $u_+$ denotes particles moving to the right, whereas $u_-$ denotes particles moving to the left. To numerically solve the entire system~(\ref{eq5}) it is convenient to simulate the equations given by Eq.(\ref{eq26}) and (\ref{eq27}). 
\begin{figure}
\includegraphics[width=8cm,origin=c,angle=0]{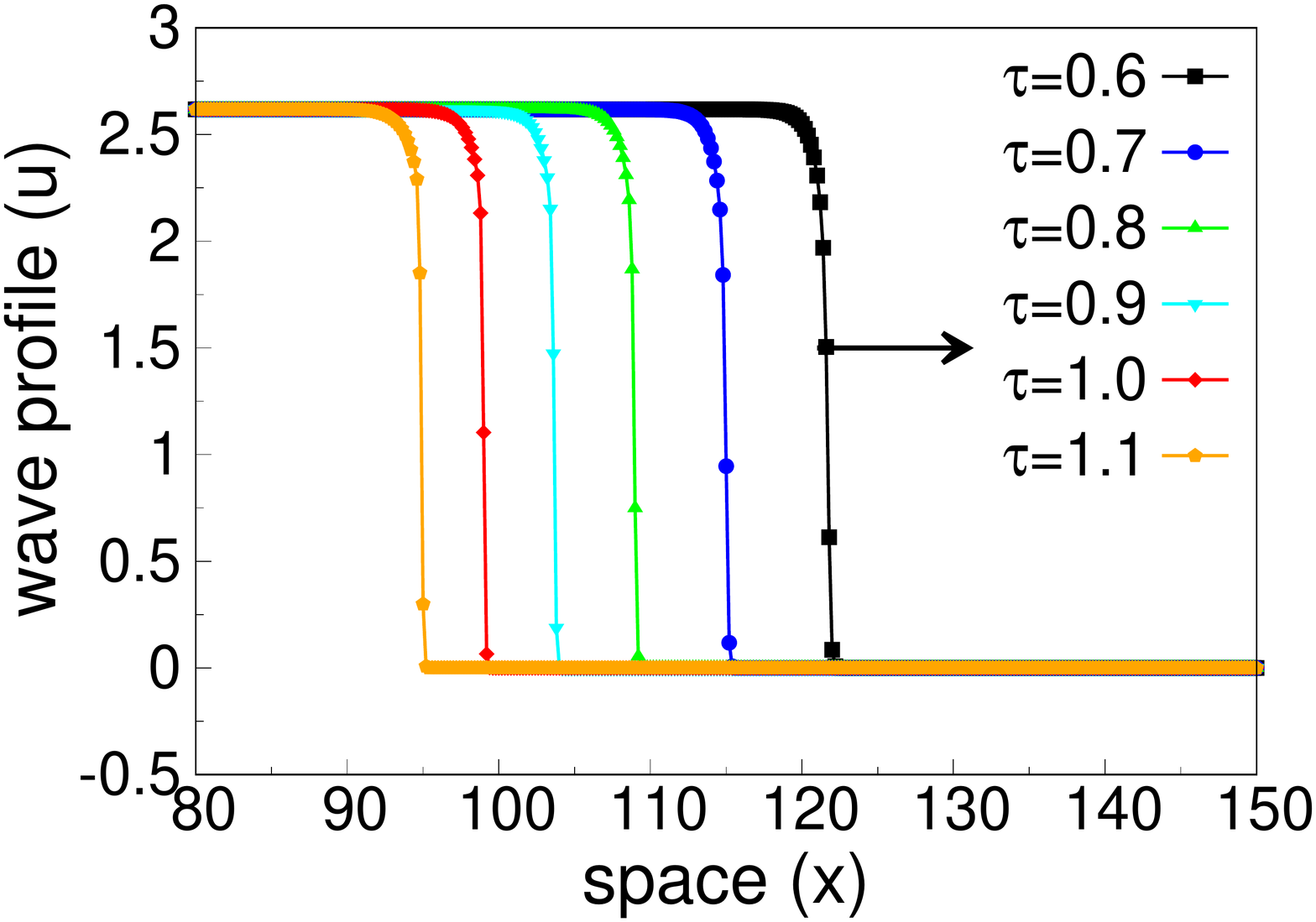}\caption{Numerically simulated density profiles of the traveling waves connecting the two steady states and moving towards the right end
for the set of parameters; $\alpha=2.612$, $\beta=0.381$, $\gamma=1.0$, $D=0.5$ and $c_{0}=1.0$ for different values of the delay time of flux:
(a) $\tau=0.6$, (b) $\tau=0.7$, (c) $\tau=0.8$, (d) $\tau=0.9$, (e) $\tau=1.0$, and (f) $\tau=1.1$ .}\label{numerical_wavefront}
\end{figure}

 A finite volume scheme is used to simulate the aforementioned system with  $f=-u^3+3u^2-u$. The numerical computations are performed in a finite space-time  domain  with a grid size, $\Delta x=0.01$ and time step, $\Delta t = 0.0005$ and maintaining Neumann boundary condition~\cite{Tilles2019}. The initial condition for $u_+$ and $u_-$ are: $u_+=u_-=2.0$ when $x\geqslant 0$ and $u_+=u_-=0$ when $x\leq 0$. The left boundary of the one dimensional spatial domain is chosen as $l=-5.0$ and the cell centre $x_i=l+(i-1)\Delta x$ and instants $t_n = n\Delta t $ constitute the discrete mesh points $(x_i ,t_n )$. The numerical solutions of Eq.(\ref{eq26}) and (\ref{eq27}) are denoted by $u_{+}^{i,n}$ and $u_{-}^{i,n}$ respectively. 
 
To obtain traveling wave solution in the hyperbolic regime, we aim to avoid numerical diffusion. Therefore, we use a nonlinear flux-limiter scheme~\cite{Logan,Sabelnikov}. The source terms are discretized following the Euler explicit method:
\begin{eqnarray}
u_{+}^{i,n+1}&=&u_{+}^{i,n}-\frac{1}{\sqrt{\tau}}\frac{\Delta t}{\Delta x}\left[u_{+}^{i,n} + \frac{\Delta x}{2}\sigma_{+}^{i,n} - u_{+}^{i-1,n} - \frac{\Delta x}{2}\sigma_{+}^{i-1,n}\right] \nonumber \\
&+& \Delta t \left[\frac{1}{2\tau}(u_{-}^{i,n}-u_{+}^{i,n})+\frac{f(u)}{2}\right] \\
u_{-}^{i,n+1}&=&u_{-}^{i,n}+\frac{1}{\sqrt{\tau}}\frac{\Delta t}{\Delta x}\left[u_{-}^{i+1,n} - \frac{\Delta x}{2}\sigma_{-}^{i+1,n} - u_{-}^{i,n} + \frac{\Delta x}{2}\sigma_{-}^{i,n}\right] \nonumber \\
&+& \Delta t \left[\frac{1}{2\tau}(u_{+}^{i,n}-u_{-}^{i,n})+\frac{f(u)}{2}\right] 
\end{eqnarray}
Here, $\sigma_{+}^{i,n}$ and $\sigma_{+}^{i,n}$ represent the nonlinear reconstruction of the slopes and obtained as:
\begin{eqnarray}
\sigma_{\pm}^{i,n}=minmod \left(\dfrac{u_{\pm}^{i,n}-u_{\pm}^{i-1,n}}{\Delta x},\dfrac{u_{\pm}^{i+1,n}-u_{\pm}^{i,n}}{\Delta x}\right)
\end{eqnarray}
where
\[
    minmod(a,b)= 
\begin{cases}
    a, & \text{if } |a|<|b|    \text{and } ab > 0\\
    b, & \text{if } |a|>|b|    \text{and } ab > 0\\
    0,              & \text{if } ab \leqslant 0.
\end{cases}
\]
We compute the system using aforementioned numerical recipe following Courant-Friedrichs-Levy (CFL) condition, i.e., $\Delta t <$ CFL$\sqrt{\tau}\Delta x$ , where CFL$ < 1$. To check the convergence of the obtained numerical solution of the simulated system to an asymptotic traveling wave, we tracked few number of points $x_c(t)$, where $u(x_p(t),t) =p$. Then, the speeds $dx_p/dt$ of these points were calculated. The values of $p$ were taken are $p$ = 0.5, 1.0, 1.5, and 2.0. The numerical solution was assumed to be converged to the asymptotic traveling wave solution when the speeds of the four level points coincided up to the fourth digit. Although, our numerical simulation can capture the entire range of $\tau>0$ unlike the limitations we find in our analytical study, we mainly are interested in the hyperbolic regime. Accordingly, the present simulations are done for $\tau > 0.502$.

\begin{figure}
\includegraphics[width=8cm,origin=c,angle=0]{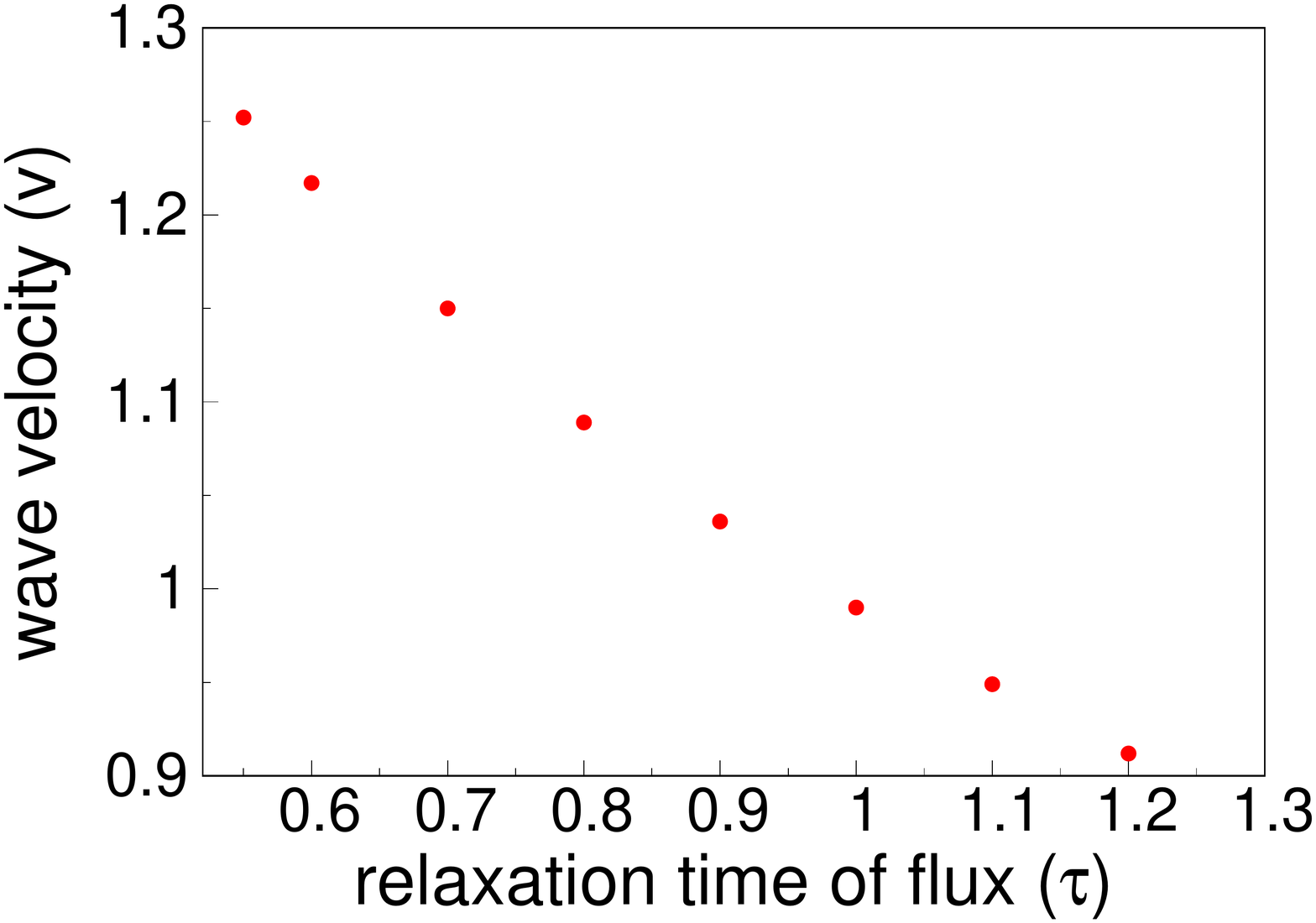}\caption{Speed of the numerically simulated traveling waves moving to the right end as a function of the delay time of flux. The values of the other  parameters are: $\alpha=2.612$, $\beta=0.381$, $\gamma=1.0$, and $D=0.5$.}\label{numerical_speed}
\end{figure}

In Fig.\ref{numerical_wavefront}, we depict the numerically simulated concentration profiles of the traveling waves. The plots are obtained at $t=100$ for  different values of $\tau$. It is apparent from the Fig.\ref{numerical_wavefront}, that the waves are moving in a positive direction i.e to the right with different speeds. In order to get a qualitative measure of the effect of $\tau$ on the modification of the speed of the traveling wave, we plot the wave speed as a function of $\tau$. The result is depicted in Fig.\ref{numerical_speed}. We observe that the wave speed decreases with the increase of $\tau$. Our numerical results are qualitatively in a good agreement with the analytical predictions.    

\section{conclusion} It is well known that linear stability
analysis is not always equipped to deal with full characterization
of a steady state. This is particularly true for steady stable
systems with strong nonlinearity. In this paper we have shown how a
non-infinitesimal perturbation acts on a homogeneous stable steady
state to grow as a traveling wave in  reaction-Cattaneo systems with cubic  nonlinearity. The nonlinear analysis has been cast in such a way that the spatiotemporal evolution of non-infinitesimal 
perturbation appears isomorphic (in form but not in content) to population dynamics or flame propagation. The result is generically distinct from the characteristic linear evolution of infinitesimal perturbation on a homogeneous steady state. Our conclusion may be summarized as follows:

(\emph{i}) In absence of any relaxation effect of the flux, the
diffusion coefficient along with the nonlinearity of the source
function determine the nature of the traveling wave solution. It is
also necessary to emphasize that the nature of the steady state
characterizes nonlinear evolution through source function. The
origin of instability therefore lies on the interplay between
diffusion and nonlinearity of the dynamics of finite perturbation.

(\emph{ii}) The nonlinear evolution of spatiotemporal perturbation
corresponding to the reaction-Cattaneo equation which describes from a
microscopic point of view a persistent random walk motion of the reacting
species having finite inertia, results in traveling wave solution modified by relaxation effect. The relaxation time reduces the wave speed when the perturbation is small and finite.

       In conclusion, when the spatiotemporal perturbations are
non-infinitesimal, the associated dynamical evolution may exhibit
various complex features. The present work is limited by
consideration of nonlinearity of cubic variety. We believe that the
approach may be extended further so that polynomial source functions
other than cubic may be exploited in physically relevant situations
for related studies in the field of active biological and soft matter.
\\

 \textbf{Acknowledgement}:  
P.G. acknowledges the funding from Department of Science and Technology, India in the form of INSPIRE Faculty Award (Grant No: IFA15/CH-201). Thanks  are due to  Department of Science and Technology, India, for a J. C. Bose National Fellowship (D.S.R.) under  Grant  No.  SB/S2/JCB-030/2015  for  partial financial support.

\bibliography{ms}
\end{document}